# Applying least absolute deviation regression to regression-type estimation of the index of a stable distribution using the characteristic function

**J. MARTIN VAN ZYL**

Department of Mathematical Statistics and Actuarial Science, University of the Free State, PO Box 339, Bloemfontein, South Africa

**Abstract** Least absolute deviation regression is applied using a fixed number of points for all values of the index to estimate the index and scale parameter of the stable distribution using regression methods based on the empirical characteristic function. The recognized fixed number of points estimation procedure uses ten points in the interval zero to one, and least squares estimation. It is shown that using the more robust least absolute regression based on iteratively re-weighted least squares outperforms the least squares procedure with respect to bias and also mean square error in smaller samples.

**Mathematical Subject** Classification 62F10; 60E07; 62E10;

## 1. Introduction

A procedure using robust LAD regression based on the empirical characteristic function (c.f.) evaluated at a fixed number of points to estimate parameters of the symmetric stable distribution is proposed. Denote the c.f. by $\phi(t)$. Let $x_1,...,x_n$ denote a sample of size n i.i.d. observations. The sample c.f. is estimated for a given value of t as $\hat{\phi}_n(t) = \frac{1}{n}\sum_{j=1}^{n} e^{itx_j}$. Suppose the c.f. is estimated at K points $t_1,...,t_K$, $k = 1,...,K$. Koutrouvelis (1980) showed that the transformation $\log(-\log(|\hat{\phi}_n(t_k)|^2))$ can be used to construct linear regression equations to estimate the parameters. The resulting regression equation is highly heteroscedastic and there is also a much more complicated autocorrelation structure than a simple autoregressive process of low order. Koutrouvelis (1980) found that the optimal values for calculating the empirical c.f. are $t_1,...,t_K, t_k = \pi k/25, k = 1,...,K$. A complication is that K depends on the unknown value of the index of the stable distribution and the estimation results are very sensitive to the number of points. Thus the two problems when using the empirical c.f. regression approach is to find the optimal value of K and which points must be chosen.

Methods were derived using a fixed number of points and other using a number which is a function of the unknown parameters. This work will focus on comparing the LAD regression estimation procedure using a fixed number of points for all values of the unknown parameters with that of Kogon and Williams (1998), which also make use of a fixed number of points.

The following aspects will be taken into account in this work:

- The interval where the residual variance of the regression reaches a minimum and is most constant. It was found that this interval is approximately for $t \in [0.5, 1.0]$ and using points chosen in this interval leads to excellent results with respect to the bias of the estimated parameters but performs reasonable with respect to MSE.

- The more robust least absolute regression (LAD) making use of iteratively reweighted least squares (IRLS) is tested. IRLS for LAD estimation use weigths with are inversely proportional to the absolute value of the residuals and may perform good in regression problems where heteroscedasticity is present. It was found that if a fixed number of points K=20, in the interval [0.1, 1.0] is used, then excellent estimation results with respect to both MSE and bias were found over the whole range of parameters.

- The sensitivity of the various procedures with respect to the number of points and which points are chosen. It was found that the LAD procedure with points chosen on the interval [0.1, 1.0], the results is robust with respect to the choice of K, but for the points $t_1,...,t_K, t_k = \pi k/25, k=1,...,K$, the wrong choice of K can lead to very biased estimators of the index when least squares estimation is performed.

The more robust LAD regression using the IRLS method and a fixed number of points outperforms the Kogon and Williams (1998) procedure in samples with up to a few hundred observations. Kogon and Williams (1998) suggested ten points [0.1, 0.2, …,1.0] using least squares regression. Some skewness in the data does have a small influence on the estimation results.

LAD estimation outperforms the Koutrouvelis (1980) estimation method where the number and choice of points where the c.f. is calculated is chosen using initial estimation of the parameters. For a given sample, the c.f., $\phi(t)$, regression equations are formed based on calculating the empirical characteristic function at points $t_1,...,t_K, t_k = \pi k/25, k=1,...,K$. In practice the parameters and specifically the index is unknown and K is a function of the unknown index of the stable distribution. When the c.f. is calculated at the optimal points this method performs excellent, but the method of Koutrouvelis (1980) is very biased when choosing K incorrectly.

This work will focus on the estimation of the index of symmetrically stable distributed data. Such data are often used in market risk analysis and especially when working with log returns of assets traded in a market. Some of this is reviewed in the books by Cizek, Härdle and Weron, eds. (2011), Gentle, Härdle, Mori, eds, 2004.

Denote the c.f. of the stable distribution by $\phi(t)$ where

$$\log\phi(t) = -\sigma^\alpha |t|^\alpha \{1 - i\beta \, sign(t)\tan(\pi\alpha/2)\} + i\mu t, \ \alpha \neq 1,$$
and $\quad \log\phi(t) = -\sigma|t|\{1 + i\beta \, sign(t)(2/\pi)\log(|t|)\} + i\mu t, \ \alpha = 1.$

The parameters are the index $\alpha \in (0,2]$, scale parameter $\sigma > 0$, coefficient of skewness $\beta \in [-1,1]$ and mode $\mu$. The symmetric case with $\mu = 0, \beta = 0$ will be considered in this work. Koutrouvelis (1980) made use of the properties of the c.f. and using the fact that $|\phi(t)|^2 = \exp(-2\sigma^\alpha |t|^\alpha)$ derived the model which does not involve $\beta$ and $\mu$ when estimating the index $\alpha$ and $\sigma$:

$$\log(-\log(|\phi(t)|^2)) = \log(2\sigma^\alpha) + \alpha \log(|t|), \qquad (1.1)$$

a simple linear regression model can be formed

$$y_k = m + \alpha \omega_k + \varepsilon_k. \qquad (1.2)$$

The c.f. is estimated for a given value of t, for a sample of size n i.i.d. observations $x_1,..., x_n$, as $\hat{\phi}_n(t) = \frac{1}{n}\sum_{j=1}^{n} e^{itx_j}$, and $y_k = \log(-\log(|\hat{\phi}_n(t_k)|^2))$, $m = \log(2\sigma^\alpha)$, $\omega_k = \log(|t_k|)$, $\varepsilon_k$ an error term. Koutrouvelis (1980) derived the optimal points $t_k = \pi k / 25, k = 1,..., K$, and optimal values of K was suggested for various sample sizes and $\alpha's$. In practice for a specific sample size $\alpha$ is unknown and choosing it incorrectly leads to incorrect estimation results.

An expression for the covariance $\text{cov}(|\hat{\phi}_n(t_j)|^2, |\hat{\phi}_n(t_k)|^2)$ and thus the variance of $|\hat{\phi}_n(t_j)|^2$ is given by Koutrouvelis (1980). This expression depends on the unknown parameters, and thus also $Var(\log(-\log(|\hat{\phi}_n(t_j)|^2)))$ making weighted regression problematic.

Paulson *et al.* (1997) showed that by using standardized data estimation results can be improved and all estimation in this work will be performed on standardized data. Koutrouvelis (1980) found that this regression equation does not depend on the location parameter. Koutrouvelis (1980) suggested using a truncated mean of 25% and the Fama and Roll ( 1971) estimator of the scale parameter $\sigma$:

$$\hat{\sigma} = (x_{.72} - x_{.28})/1.654,$$

where the .72 and .28 denote percentiles of the data, to standardize the data. The same standardization will be used before applying the LAD regression. Kogon and Williams (1998) used initial estimators using the method of McCullogh (1986) to perform standardization.

The LAD results will be compared on standardized data as described above against the Koutrouvelis method assuming K known, K estimated and also the Kogon-Williams procedure (Borak and Weron (2010c)). The optimal K for the procedure of Koutrouvelis (1980) is chosen using an initial estimate of the index using the Mcullogh (1986) estimate as applied by Borak and Weron (1910a).

Much research was done on using an approximate covariance matrix and generalized least squares to estimate the parameters. An excellent overview of this approach is given by Besbeas and Morgan (2008). They suggested using arithmetic spacing of t's which performs very well but the optimal number of points chosen is also not independent of the unknown parameters. The work of Feuerverger and McDunnough (1981a, 1981b, 1981c), Koutrouvelis and Bauer (1982) is also of importance where weighted least squares and generalised least squares were applied. But also no definite number of sampling points which will perform good over the whole range of the index.

It is shown in section 2 that the residual variance is highly heteroscedastistic with respect to t. This might lead to a decrease in the efficiency, and also incorrect estimates of the variances of the estimated parameters. The variance of the residuals, $\varepsilon's$ for a given t and the true parameters, is estimated using simulated samples. Residuals of a sample was calculated using the true parameters as $\varepsilon_k = y_k - m - \alpha \omega_k$, and from these residuals the variance, $Var(\varepsilon_k) = Var(\varepsilon_k | t, \sigma, \alpha)$, was estimated. The sensitivity of the Koutrouvelis procedure with respect to K is also investigated. In section 3 simulation studies were conducted to compare the various estimators.

## 2. The residual variance when using regression type methods based on the empirical characteristic function

The residual variance, autocorrelation structure when using least squares and the sensitivity when choosing K incorrectly with respect to the optimal K will be investigated in this section. The programs of Borak and Weron (2010a), (2010b), (2010c), Borak *et al* (2011) were used when applying the method of Kogon and Williams (1998) and also when applying the method of Koutrouvelis (1980) using initial estimated parameters. Their initial estimation and also standardization was according to the work of McCullogh (1986).

In figure 1 the estimated residual variance based on the true parameters using (1.2) and non-standardized data is plotted for various values of t. The variance of the residuals of $M = 1000$ samples of size $n = 200$ each, with respect to t is shown. The data was simulated with $\alpha = 1.5$, $\sigma = 0.1$, $\beta = 0$, $\mu = 0$. Similar patterns was observed for other values of $\alpha, \sigma$. The error variance is smallest in the interval with t between 0.5 and 1.0. The true parameters are used in the calculation of the error variance, for example

$\hat{\sigma}_t^2 = \hat{\sigma}_t^2 \mid t, \sigma, \alpha = \text{var}(\varepsilon_1, ..., \varepsilon_M \mid t, \alpha, \sigma)$, , where $\varepsilon_j \mid \alpha, m, t = y_j - m - \alpha \log(t)$, $y_j = \log(-\log(\mid \hat{\phi}_n(t) \mid^2))$, $j = 1, ..., M$ .

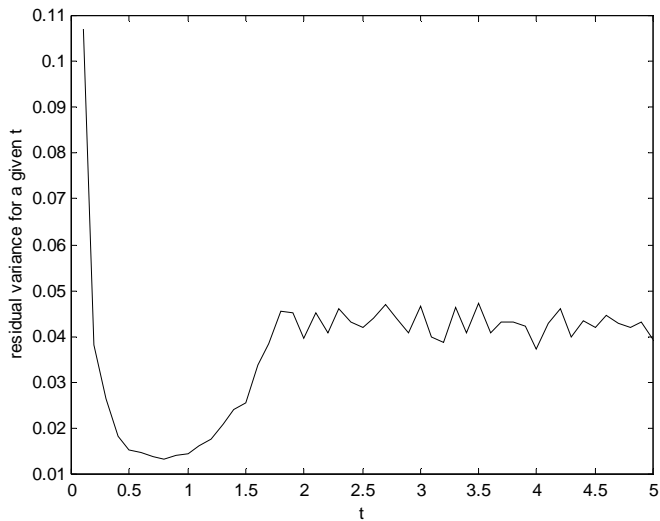

**Figure 1** Estimated residual variances for given values of t, $\alpha = 1.5, \sigma = 0.1, n = 200$.

In figure 2 the estimated residual variance based on the estimated parameters using (1.2) and standardized data is plotted for various values of t. The data standardized using the program by Borak and Weron (2010a). Similar patterns of the error variances were observed for different values of the index. The autocorrelation function plot of the residuals for a specific sample is shown in figure 3, which shows that a simple autoregressive type model will not fit the residuals. Experimentation showed that the pattern is a complicated ARMA type model with terms of high order which is not easily identifiable, making the use of regression models with autocorrelated residuals very difficult.

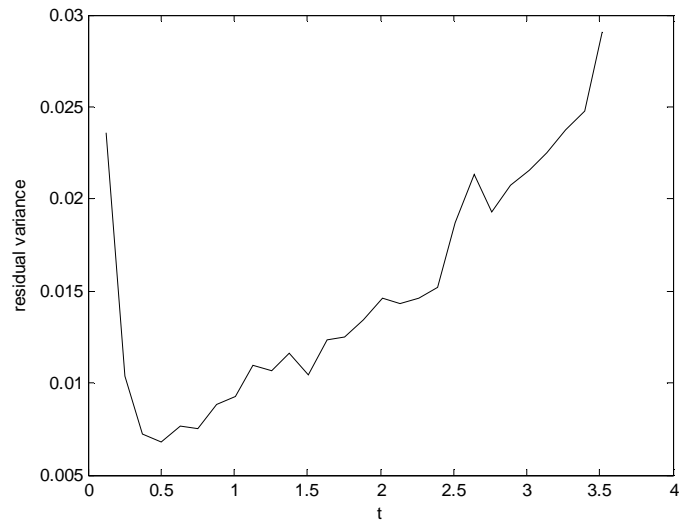

**Figure 2** Estimated residual variances for K=28 values of t, $\alpha = 0.9, \sigma = 1.0, n = 200$. Standardized data and variances calculated from 500 regressions.

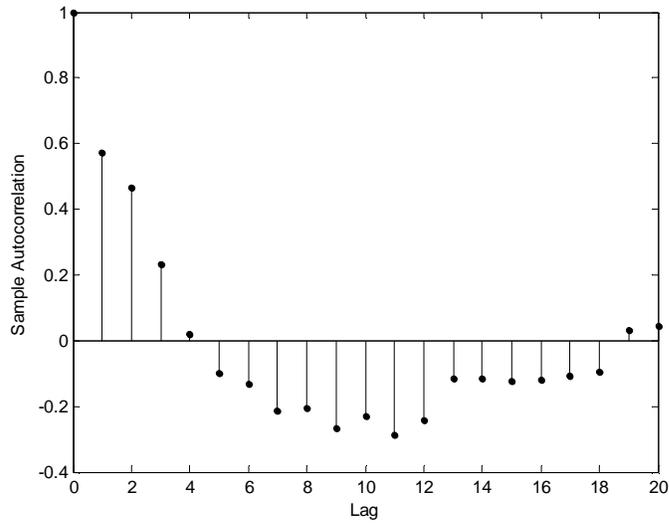

**Figure 3** ACF of residuals, $\alpha = 0.9, \sigma = 1.0, n = 200$. Standardized data.

The bias when using the incorrect number of points when using the Koutrouvelis (1980) procedure on standardized data is shown. The estimated parameters was calculated based on 500 estimated values of the index $\alpha = 1.3$ for which the optimal K=22. The LAD estimated index for various values of K is also shown, and it can be seen that the result is not very sensitive with respect to K.

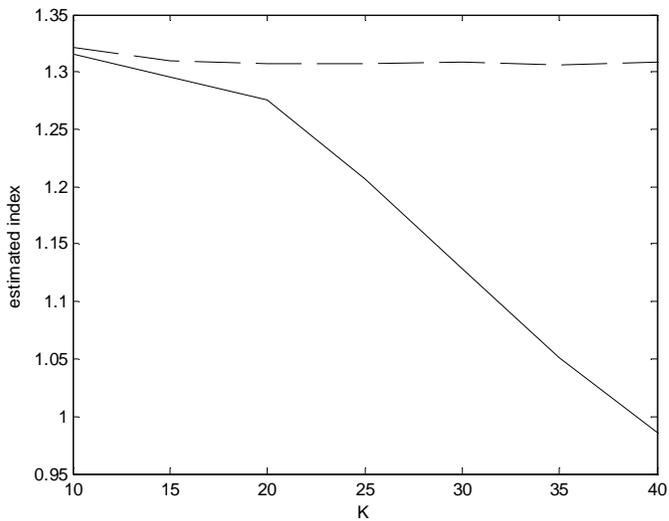

**Figure 4** Estimated index, various values of K, $\alpha = 1.3, \sigma = 1.0, \beta = 0.0, n = 200$. The solid line when using $\pi k / 25, k = 1, ..., K$. Average of 500 estimated indexes at each K. and the dashed line K LAD regression using K points in the interval [0.1:1.0].

This small error variance together with the research of Kogon and Williams (1998) motivated the use of interval [0.1, 1.0] when using LAD regression. Experimentation showed that 20 uniformly distributed points yielded good estimation results when using LAD, with respect to bias and MSE.

## 3. Comparison between estimation procedures

In this section a simulation study was conducted. Standardization was performed on all the data before estimation and the location parameter $\mu = 0$, scale parameter $\sigma = 1.0$ was used for all the simulations. Samples from symmetric distributions with location parameter zero were considered.

For the LAD method, data was standardized using the Fama and Roll (1971) estimator and a 25% trimmed mean. K=20 points were used for all values of the index, and the points used were $t_k = 0.1 + 0.05(k-1), k = 1,...,20$.

Iteratively reweighted least squares (IRLS) can be used to minimize with respect to the absolute value norm if the weight matrix is a diagonal matrix with diagonal elements equal to the inverse of the absolute value of the residuals. This technique was applied to find the LAD estimators. The method involves a weighted least squares multiple linear regression form, calculated iteratively, using a diagonal weight matrix $W^{(j)}$ with diagonal elements with $w_i^{(j)} = 1/(1+|e_i|), i = 1,...,20$, where $e_i$ is the i-th residual at iteration j. This form is chosen to avoid division by zero and it can be see that it is weighted regression with weights inversely related to the size of the absolute value of the residuals. The results are based on 10000 simulated samples each time. No adjustment was made if the LAD estimate was larger than 2.

For the Kogon and Williams (1998) method the points $t_k = 0.1 + 0.1(k-1), k = 1,...,10$ were used. The results of the Koutrouvelis (1980) procedure choosing K using the true index is included as a reference. It should be noted this is not comparable to the other three methods with respect to practical problems where there is no prior knowledge of the parameters and all are parameters are calculated using the sample. The Borak and Weron (2010a) method base the number of points on estimated initial values.

### 3.1 Results for the estimation of the index

The assumption is not made that the mean is zero and standardization is carried out also with respect to the mean in the simulations. For the index estimation, the LAD methods outperforms the other procedures with respect to MSE in smaller samples and much the same when it was tested on skewed data where $\beta = 0.5$. It seems that in larger samples finding initial estimates and then using the Koutrouvelis optimal points yields the best results.

It can be noted that the Fama and Roll (1971) estimation method which is used for standardization, is strictly speaking valid for $\alpha \geq 1$, which may explain the weak estimation results when using these estimates to standardize the data when $\alpha < 1$, and thus leads to poor results especially when $\alpha = 0.7$ and using the method of Koutrouvelis (1980). This point was not included in the figures.

In table 1 the performance of using least squares estimation using ten points uniformly chosen in the interval [0.5,1] is shown. It can be seen that it yields excellent results with respect to bias, but the MSE is weaker than the Kogon-Williams procedure. Overall LAD regression is best to use when both bias and MSE is taken into account with a MSE considered as least or more important than bias.

| $\alpha$ | LAD | | | Least Squares [0.5,1.0] | | | Kogon-Williams [0.1,1.0] | | |
|---|---|---|---|---|---|---|---|---|---|
| | $\hat{\alpha}$ | Bias | MSE | $\hat{\alpha}$ | Bias | MSE | $\hat{\alpha}$ | Bias | MSE |
| 1.9 | 1.9043 | 0.0043 | 0.0069 | 1.8934 | -0.0066 | 0.0101 | 1.9046 | 0.0046 | 0.0073 |
| 1.5 | 1.5103 | 0.0103 | 0.0159 | 1.4947 | -0.0053 | 0.0271 | 1.5131 | 0.0131 | 0.0175 |
| 1.3 | 1.3087 | 0.0087 | 0.0141 | 1.2958 | -0.0042 | 0.0329 | 1.3110 | 0.0110 | 0.0150 |
| 1.1 | 1.1070 | 0.0070 | 0.0122 | 1.0971 | -0.0029 | 0.0387 | 1.1100 | 0.0100 | 0.0127 |
| 0.9 | 0.9034 | 0.0034 | 0.0094 | 0.8978 | -0.0022 | 0.0420 | 0.9065 | 0.0065 | 0.0095 |
| 0.7 | 0.7035 | 0.0035 | 0.0075 | 0.6990 | -0.0010 | 0.0440 | 0.7070 | 0.0070 | 0.0075 |

**Table 1** Comparison of estimation procedures of $\alpha$ with respect to bias and MSE,

$\sigma = 1.0, \mu = 0, \beta = 0, n = 100$. LAD using 20 points on the interval [0.1,1.0], least squares 10 points on the interval [0.5,1], and the Kogon-Williams procedure 10 points on the interval [0.1,1].

| $\alpha$ | LAD | | | Koutrouvelis (K optimal \|true $\alpha$) | | | Koutrouvelis (Borak and Weron) | | | Kogon-Williams | | |
|---|---|---|---|---|---|---|---|---|---|---|---|---|
| | $\hat{\alpha}$ | Bias | MSE | $\hat{\alpha}$ | Bias | MSE | $\hat{\alpha}$ | Bias | MSE | $\hat{\alpha}$ | Bias | MSE |
| 1.9 | 1.9053 | 0.0053 | 0.0125 | 1.8965 | -0.0035 | 0.0133 | 1.8921 | -0.0079 | 0.0121 | 1.9039 | 0.0039 | 0.0121 |
| 1.5 | 1.5230 | 0.0230 | 0.0320 | 1.5151 | 0.0151 | 0.0287 | 1.4986 | -0.0014 | 0.0369 | 1.5283 | 0.0283 | 0.0347 |
| 1.3 | 1.3185 | 0.0185 | 0.0299 | 1.3129 | 0.0129 | 0.0269 | 1.2807 | -0.0193 | 0.0344 | 1.3245 | 0.0245 | 0.0319 |
| 1.1 | 1.1120 | 0.0120 | 0.0250 | 1.1090 | 0.0090 | 0.0229 | 1.0692 | -0.0308 | 0.0278 | 1.1181 | 0.0181 | 0.0258 |
| 0.9 | 0.9073 | 0.0073 | 0.0203 | 0.9059 | 0.0059 | 0.0188 | 0.8538 | -0.0462 | 0.0310 | 0.9160 | 0.0160 | 0.0207 |
| 0.7 | 0.7059 | 0.0059 | 0.0156 | 0.7049 | 0.0049 | 0.0148 | 0.5767 | -0.1233 | 0.0416 | 0.7128 | 0.0128 | 0.0158 |

**Table 2** Comparison of estimation procedures of $\alpha$ with respect to bias and MSE, $\sigma = 1.0, \mu = 0, \beta = 0, n = 100$.

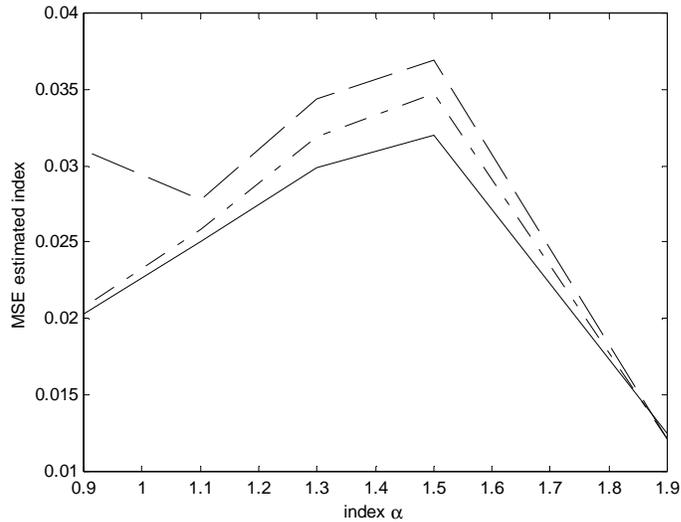

**Figure 5** MSE of three procedures for various values of the index $\alpha$. Symmetric data and estimation performed on standardized data, n=100, 10000 estimated samples. Solid line - LAD, dash dot - Kogon-Williams method and dashed the method of Koutrouvelis with the number of points chosen using the method of McCullogh.

| $\alpha$ | LAD | | | Koutrouvelis (K optimal \|true $\alpha$) | | | Koutrouvelis (Borak and Weron) | | | Kogon-Williams | | |
|---|---|---|---|---|---|---|---|---|---|---|---|---|
| | $\hat{\alpha}$ | Bias | MSE | $\hat{\alpha}$ | Bias | MSE | $\hat{\alpha}$ | Bias | MSE | $\hat{\alpha}$ | Bias | MSE |
| 1.9 | 1.9033 | 0.0033 | 0.0069 | 1.9010 | 0.0010 | 0.0070 | 1.8981 | -0.0019 | 0.0065 | 1.9034 | 0.0034 | 0.0073 |
| 1.5 | 1.5104 | 0.0104 | 0.0154 | 1.5038 | 0.0038 | 0.0132 | 1.5001 | 0.0001 | 0.0167 | 1.5135 | 0.0135 | 0.0169 |
| 1.3 | 1.3093 | 0.0093 | 0.0143 | 1.2542 | -0.0458 | 0.0139 | 1.2916 | -0.0084 | 0.0152 | 1.3123 | 0.0123 | 0.0153 |
| 1.1 | 1.1071 | 0.0071 | 0.0119 | 1.0687 | -0.0313 | 0.0104 | 1.0846 | -0.0154 | 0.0121 | 1.1095 | 0.0095 | 0.0123 |
| 0.9 | 0.9039 | 0.0039 | 0.0096 | 0.8830 | -0.0170 | 0.0074 | 0.8914 | -0.0086 | 0.0097 | 0.9069 | 0.0069 | 0.0097 |
| 0.7 | 0.7031 | 0.0031 | 0.0075 | 0.7002 | 0.0002 | 0.0054 | 0.6340 | -0.0660 | 0.0171 | 0.7062 | 0.0062 | 0.0074 |

**Table 3** Comparison of estimation procedures of $\alpha$ with respect to bias and MSE, $\sigma = 1.0, \mu = 0, \beta = 0, n = 200$.

The MSE of the various procedures is shown in figure 6.

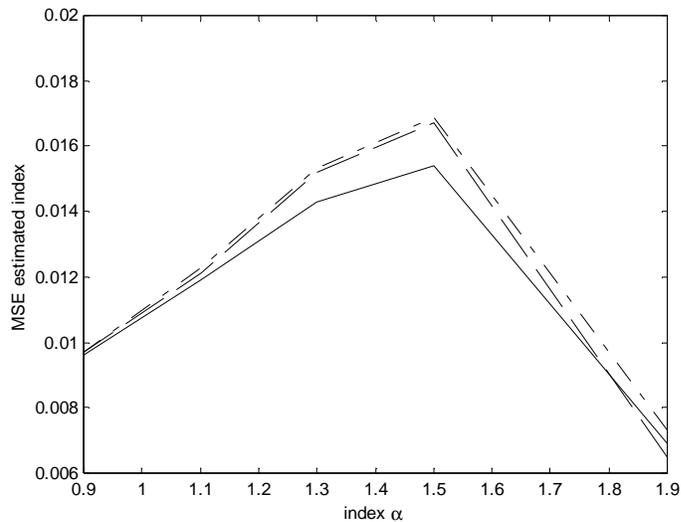

**Figure 6** MSE of three procedures for various values of the index $\alpha$. Symmetric data and estimation performed on standardized data, n=200, 10000 estimated samples. Solid line - LAD, dash dot - Kogon-Williams method and dashed the method of Koutrouvelis with the number of points chosen using the method of McCullogh.

| $\alpha$ | LAD | | | Koutrouvelis (K optimal \|true $\alpha$) | | | Koutrouvelis | | | Kogon-Williams | | |
|---|---|---|---|---|---|---|---|---|---|---|---|---|
| | $\hat{\alpha}$ | Bias | MSE | $\hat{\alpha}$ | Bias | MSE | $\hat{\alpha}$ | Bias | MSE | $\hat{\alpha}$ | Bias | MSE |
| 1.9 | 1.9007 | 0.0007 | 0.0019 | 1.9002 | 0.0002 | 0.0019 | 1.8999 | -0.0001 | 0.0019 | 1.9010 | 0.0010 | 0.0022 |
| 1.5 | 1.5035 | 0.0035 | 0.0039 | 1.5020 | 0.0020 | 0.0033 | 1.5034 | 0.0034 | 0.0037 | 1.5041 | 0.0041 | 0.0043 |
| 1.3 | 1.3024 | 0.0024 | 0.0035 | 1.2987 | -0.0013 | 0.0031 | 1.2970 | -0.0030 | 0.0034 | 1.3030 | 0.0030 | 0.0038 |
| 1.1 | 1.1014 | 0.0014 | 0.0030 | 1.0990 | -0.0010 | 0.0025 | 1.0969 | -0.0031 | 0.0030 | 1.1019 | 0.0019 | 0.0031 |
| 0.9 | 0.9015 | 0.0015 | 0.0024 | 0.9001 | 0.0001 | 0.0021 | 0.8983 | -0.0017 | 0.0022 | 0.9022 | 0.0022 | 0.0024 |
| 0.7 | 0.7005 | 0.0005 | 0.0018 | 0.7010 | 0.0010 | 0.0014 | 0.6794 | -0.0206 | 0.0034 | 0.7014 | 0.0014 | 0.0018 |

**Table 4** Comparison of estimation procedures of $\alpha$ with respect to bias and MSE, $\sigma = 1.0, \mu = 0, \beta = 0, n = 800$.

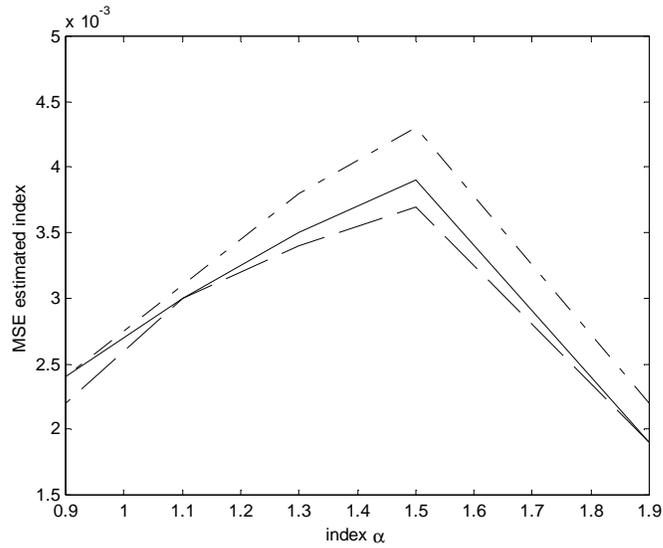

**Figure 7** MSE of three procedures for various values of the index $\alpha$. Symmetric data and estimation performed on standardized data, n=800, 10000 estimated samples. Solid line - LAD, dash dot - Kogon-Williams method and dashed the method of Koutrouvelis with the number of points chosen using the method of McCullogh.

| $\alpha$ | LAD | | | Koutrouvelis (K optimal \|true $\alpha$) | | | Koutrouvelis | | | Kogon-Williams | | |
|---|---|---|---|---|---|---|---|---|---|---|---|---|
| | $\hat{\alpha}$ | Bias | MSE | $\hat{\alpha}$ | Bias | MSE | $\hat{\alpha}$ | Bias | MSE | $\hat{\alpha}$ | Bias | MSE |
| 1.9 | 1.9034 | 0.0034 | 0.0070 | 1.8960 | -0.0040 | 0.0074 | 1.8982 | -0.0018 | 0.0065 | 1.9034 | 0.0034 | 0.0073 |
| 1.5 | 1.5119 | 0.0119 | 0.0159 | 1.5058 | 0.0058 | 0.0134 | 1.5021 | 0.0021 | 0.0162 | 1.5141 | 0.0141 | 0.0170 |
| 1.3 | 1.3102 | 0.0102 | 0.0151 | 1.2667 | -0.0333 | 0.0119 | 1.2909 | -0.0091 | 0.0154 | 1.3123 | 0.0123 | 0.0155 |
| 1.1 | 1.1088 | 0.0088 | 0.0129 | 1.0879 | -0.0121 | 0.0086 | 1.0875 | -0.0125 | 0.0123 | 1.1135 | 0.0135 | 0.0125 |
| 0.9 | 0.9078 | 0.0078 | 0.0107 | 0.8949 | -0.0051 | 0.0064 | 0.8853 | -0.0147 | 0.0107 | 0.9157 | 0.0157 | 0.0100 |
| 0.7 | 0.7066 | 0.0066 | 0.0085 | 0.7031 | 0.0031 | 0.0048 | 0.6080 | -0.0920 | 0.0226 | 0.7164 | 0.0164 | 0.0081 |

**Table 5** Comparison of estimation procedures of $\alpha$ with respect to bias and MSE, $\sigma = 1.0, \mu = 0, \beta = 0.5, n = 200$.

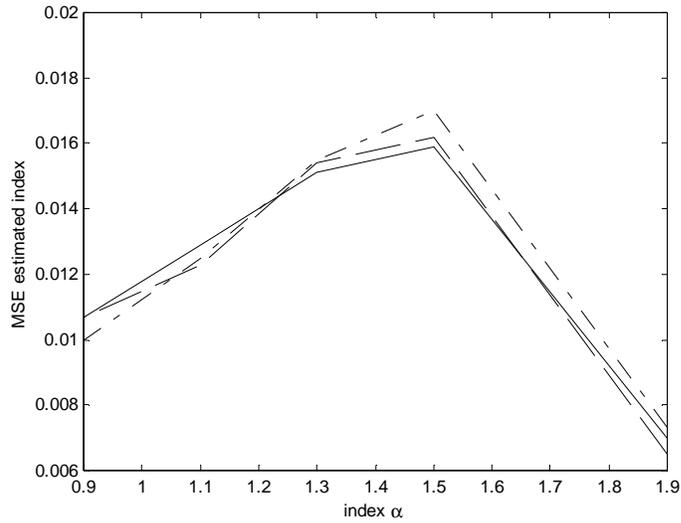

**Figure 8** MSE of three procedures for various values of the index $\alpha$. $\beta = 0.5$ and estimation performed on standardized data, n=200, 10000 estimated samples. Solid line - LAD, dash dot - Kogon-Williams method and dashed the method of Koutrouvelis with the number of points chosen using the method of McCullogh.

### 3.2 Results for the estimation of the scale parameter

There is not much difference between the estimators of the scale parameters. Again it seems that LAD will be best to apply when the samples are smaller and in large samples the other procedures performs well.

|  | LAD | | | Koutrouvelis (K optimal) | | | Koutrouvelis | | | Kogon-Williams | | |
|---|---|---|---|---|---|---|---|---|---|---|---|---|
|  | $\hat{\sigma}$ | Bias | MSE | $\hat{\sigma}$ | Bias | MSE | $\hat{\sigma}$ | Bias | MSE | $\hat{\sigma}$ | Bias | MSE |
| 1.9 | 0.9914 | -0.0086 | 0.0073 | 0.9867 | -0.0133 | 0.0073 | 0.9877 | -0.0123 | 0.0121 | 0.9920 | -0.0080 | 0.0073 |
| 1.5 | 0.9942 | -0.0058 | 0.0135 | 0.9891 | -0.0109 | 0.0139 | 0.9806 | -0.0194 | 0.0369 | 0.9954 | -0.0046 | 0.0139 |
| 1.3 | 0.9917 | -0.0083 | 0.0177 | 0.9875 | -0.0125 | 0.0178 | 0.9695 | -0.0305 | 0.0344 | 0.9940 | -0.0060 | 0.0178 |
| 1.1 | 0.9868 | -0.0132 | 0.0249 | 0.9840 | -0.0160 | 0.0248 | 0.9660 | -0.0340 | 0.0278 | 0.9915 | -0.0085 | 0.0248 |
| 0.9 | 0.9869 | -0.0131 | 0.0375 | 0.9854 | -0.0146 | 0.0367 | 0.9763 | -0.0237 | 0.0310 | 0.9952 | -0.0048 | 0.0367 |
| 0.7 | 0.9903 | -0.0097 | 0.0640 | 0.9888 | -0.0112 | 0.0600 | 1.3340 | 0.3340 | 0.0416 | 0.9987 | -0.0013 | 0.0600 |

**Table 6** Comparison of estimation procedures of $\sigma$ with respect to bias and MSE, $\sigma = 1.0, \mu = 0, \beta = 0, n = 100$.

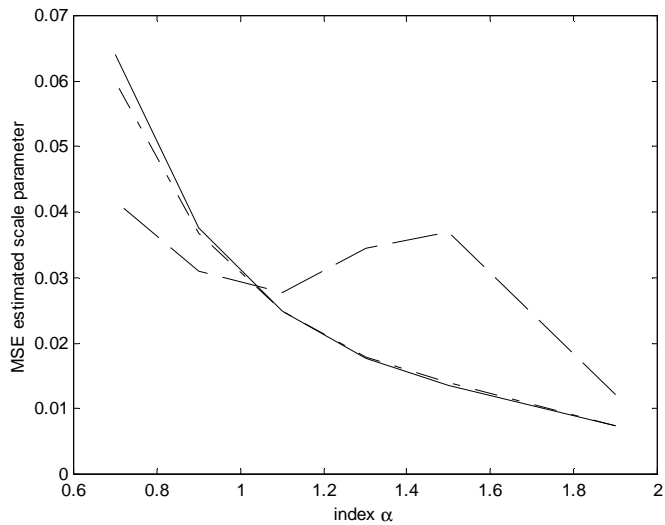

**Figure 9** MSE of three procedures for various values of the scale parameter $\sigma$. Symmetric data and estimation performed on standardized data, n=100, 10000 samples. Solid line - LAD, dash dot - Kogon-Williams method and dashed the method of Koutrouvelis with the number of points chosen using the method of McCullogh.

|     | LAD | | | Koutrouvelis (K optimal) | | | Koutrouvelis | | | Kogon-Williams | | |
| --- | --- | --- | --- | --- | --- | --- | --- | --- | --- | --- | --- | --- |
|     | $\hat{\sigma}$ | Bias | MSE | $\hat{\sigma}$ | Bias | MSE | $\hat{\sigma}$ | Bias | MSE | $\hat{\sigma}$ | Bias | MSE |
| 1.9 | 0.9954 | -0.0046 | 0.0036 | 0.9943 | -0.0057 | 0.0036 | 0.9939 | -0.0061 | 0.0065 | 0.9957 | -0.0043 | 0.0036 |
| 1.5 | 0.9958 | -0.0042 | 0.0066 | 0.9918 | -0.0082 | 0.0068 | 0.9900 | -0.0100 | 0.0167 | 0.9965 | -0.0035 | 0.0068 |
| 1.3 | 0.9968 | -0.0032 | 0.0090 | 0.9668 | -0.0332 | 0.0092 | 0.9865 | -0.0135 | 0.0152 | 0.9979 | -0.0021 | 0.0092 |
| 1.1 | 0.9957 | -0.0043 | 0.0122 | 0.9727 | -0.0273 | 0.0123 | 0.9816 | -0.0184 | 0.0121 | 0.9977 | -0.0023 | 0.0123 |
| 0.9 | 0.9926 | -0.0074 | 0.0182 | 0.9809 | -0.0191 | 0.0181 | 0.9875 | -0.0125 | 0.0097 | 0.9958 | -0.0042 | 0.0181 |
| 0.7 | 0.9961 | -0.0039 | 0.0307 | 0.9930 | -0.0070 | 0.0296 | 0.9967 | -0.0033 | 0.0171 | 0.9999 | -0.0001 | 0.0296 |

**Table 7** Comparison of estimation procedures of $\sigma$ with respect to bias and MSE, $\sigma = 1.0, \mu = 0, \beta = 0, n = 200$.

A plot of the bias of the various procedures is shown in figure 6. It can be seen that overall all the methods performs well with respect to the estimation of the scale parameter and there is little difference between the methods.

In the following tables the procedures were compared using skewed data with $\beta = 0.5$. It can be seen that using LAD yielded good results.

|     | LAD | | | Koutrouvelis (K optimal) | | | Koutrouvelis | | | Kogon-Williams | | |
| --- | --- | --- | --- | --- | --- | --- | --- | --- | --- | --- | --- | --- |
|     | $\hat{\sigma}$ | Bias | MSE | $\hat{\sigma}$ | Bias | MSE | $\hat{\sigma}$ | Bias | MSE | $\hat{\sigma}$ | Bias | MSE |
| 1.9 | 0.9960 | -0.0040 | 0.0037 | 0.9948 | -0.0052 | 0.0037 | 0.9943 | -0.0057 | 0.0066 | 0.9963 | -0.0037 | 0.0037 |
| 1.5 | 0.9963 | -0.0037 | 0.0067 | 0.9929 | -0.0071 | 0.0068 | 0.9912 | -0.0088 | 0.0162 | 0.9966 | -0.0034 | 0.0068 |
| 1.3 | 0.9964 | -0.0036 | 0.0090 | 0.9713 | -0.0287 | 0.0087 | 0.9849 | -0.0151 | 0.0154 | 0.9977 | -0.0023 | 0.0087 |
| 1.1 | 0.9949 | -0.0051 | 0.0134 | 0.9807 | -0.0193 | 0.0123 | 0.9813 | -0.0187 | 0.0123 | 1.0000 | -0.0000 | 0.0123 |
| 0.9 | 0.9991 | -0.0009 | 0.0214 | 0.9867 | -0.0133 | 0.0182 | 0.9953 | -0.0047 | 0.0107 | 1.0079 | 0.0079 | 0.0182 |
| 0.7 | 0.9987 | -0.0013 | 0.0399 | 0.9926 | -0.0074 | 0.0294 | 0.9897 | -0.0103 | 0.0226 | 1.0112 | 0.0112 | 0.0294 |

**Table 8** Comparison of estimation procedures of $\sigma$ with respect to bias and MSE, $\sigma = 1.0, \mu = 0, \beta = 0.5, n = 200$.

In table 9 results are given for sample size n=800, 10000 simulated samples.

|     | LAD | | | Koutrouvelis (K optimal) | | | Koutrouvelis | | | Kogon-Williams | | |
| --- | --- | --- | --- | --- | --- | --- | --- | --- | --- | --- | --- | --- |
|     | $\hat{\sigma}$ | Bias | MSE | $\hat{\sigma}$ | Bias | MSE | $\hat{\sigma}$ | Bias | MSE | $\hat{\sigma}$ | Bias | MSE |
| 1.9 | 0.9995 | -0.0005 | 0.0009 | 0.9993 | -0.0007 | 0.0009 | 0.9992 | -0.0008 | 0.0019 | 0.9996 | -0.0004 | 0.0009 |
| 1.5 | 0.9995 | -0.0005 | 0.0016 | 0.9987 | -0.0013 | 0.0017 | 0.9996 | -0.0004 | 0.0037 | 0.9996 | -0.0004 | 0.0017 |
| 1.3 | 0.9983 | -0.0017 | 0.0022 | 0.9960 | -0.0040 | 0.0023 | 0.9950 | -0.0050 | 0.0034 | 0.9985 | -0.0015 | 0.0023 |
| 1.1 | 0.9982 | -0.0018 | 0.0031 | 0.9966 | -0.0034 | 0.0031 | 0.9953 | -0.0047 | 0.0030 | 0.9986 | -0.0014 | 0.0031 |
| 0.9 | 0.9989 | -0.0011 | 0.0045 | 0.9977 | -0.0023 | 0.0046 | 0.9969 | -0.0031 | 0.0022 | 0.9996 | -0.0004 | 0.0046 |
| 0.7 | 0.9972 | -0.0028 | 0.0074 | 0.9988 | -0.0012 | 0.0074 | 1.0048 | 0.0048 | 0.0034 | 0.9985 | -0.0015 | 0.0074 |

**Table 9** Comparison of estimation procedures of $\sigma$ with respect to bias and MSE, $\sigma = 1.0, \mu = 0, \beta = 0, n = 800$.

It can be seen that the regression using LAD and a fixed number of points to evaluate the empirical characteristic function performs almost everywhere better that the Kogon-Williams and the Koutrouvelis procedure where the number of points is based on initial estimated parameters.

## 4. Conclusions

LAD regression performs better than using least squares Kogon and Williams (1998). The estimated $\hat{\alpha}$ can also be used to guess the best value of K, and then performing estimation based on the points $t_k = \pi k / 25, k = 1,...,K$. Because the bias is very small, also when using least squares, using the interval [0.5,1.0], might be a very easy to calculate good estimate of K before using the optimal points derived by Koutrouvelis (1980).

Refinements can be made to this procedure, with respect to the number of t's used in the regression. The simulation study shows that the method performs well with respect to bias and MSE over the whole range of parameters commonly encountered in practical problems.